\documentclass[a4paper,twocolumn,english,aps,superscriptaddress,prl,twocolumn]{revtex4-2}
\usepackage[T1]{fontenc}
\usepackage[latin9]{inputenc}
\usepackage{color}
\usepackage{babel}
\usepackage{verbatim}
\usepackage{bm}
\usepackage{amsmath}
\usepackage{amssymb}
\usepackage{graphicx}
\usepackage{esint}
\usepackage[pdfusetitle,
 bookmarks=true,bookmarksnumbered=false,bookmarksopen=false,
 breaklinks=false,pdfborder={0 0 1},backref=false,colorlinks=false]
 {hyperref}

\makeatletter

\newcommand*\LyXZeroWidthSpace{\hspace{0pt}}

\usepackage{babel}

\makeatother

\begin{document}
\renewcommand{\figurename}{FIG.} 
\title{Hidden topology and strong quantum metric bounds in trivial systems}
\author{Chang-An Li}
\email{changanli@ustc.edu.cn}

\affiliation{Hefei National Laboratory, Hefei, Anhui 230088, China}
\affiliation{School of Emerging Technology, University of Science and Technology
of China, Hefei, Anhui 230026, China}
\author{Yulin Qin}
\affiliation{Department of Physics, Fudan University, Shanghai 200433, China}
\affiliation{Department of Physics, School of Science, Westlake University, Hangzhou
310024, Zhejiang, China}
\author{Bo Fu}
\email{fubo@gbu.edu.cn}

\affiliation{School of Sciences, Great Bay University, Dongguan, China}
\author{Jian Li}
\affiliation{Department of Physics, School of Science, Westlake University, Hangzhou
310024, Zhejiang, China}
\date{\today }
\begin{abstract}
The quantum metric integral (QMI) in two-dimensional (2D) systems
is conventionally bounded from below by the Chern number. For systems
with zero Chern number or identically vanishing Berry curvature, however,
this bound becomes trivial and provides no useful geometric constraints.
Here, we develop a dimension-reduction framework that decomposes the
2D QMI into lower-dimensional components in a nested-loop way. With
this method, we establish a nonzero lower bound on the QMI arising
from one-dimensional topological obstructions even when the conventional
2D topology is trivial. We explicitly demonstrate this mechanism in
a tilted 2D Su-Schrieffer-Heeger model and an anisotropic Wilson-Dirac
model with chiral symmetry. The resulting lower bounds of QMI are
determined by the quantized Wannier bands along two different directions.
We further investigate the quantum geometry in higher-order topological
phases following the same strategy. By introducing Wannier-band basis
obtained from the nested Wilson loop, we demonstrate that the Wannier-band
QMI is  bounded from below by the higher-order topological invariant,
e.g. the quadrupole moment in Benalcazar-Bernevig-Hughes model. Our
results  establish nonzero lower bounds on QMI from a dimension-reduction
framework, thereby generalizing the fundamental relation between quantum
geometry and topology.
\end{abstract}
\maketitle

\section{Introduction}

The geometry of quantum states in parameter space encodes rich physical
information that has become central to modern condensed matter physics.
A fundamental quantity in this framework is the quantum geometric
tensor, which characterizes how quantum states vary under infinitesimal
variations of crystal momentum $\mathbf{k}$ across the Brillouin
zone \citep{Provost1980MP,Resta11EJP}. Its real symmetric part, the
quantum metric tensor $g_{\mu\nu}(\mathbf{k})$, measures the Fubini-Study
distance between neighboring Bloch states and provides a notion of
quantum distance that is gauge-invariant and physically meaningful.
Its imaginary antisymmetric part is the Berry curvature $F_{\mu\nu}(\mathbf{k})$,
whose integral over the Brillouin zone (BZ) yields the Chern number.
While the Berry curvature has long been recognized vital to the classification
of topological phases \citep{XiaoD10rmp}, the quantum metric has
recently emerged as an equally important geometric object with broad
physical consequences \citep{Torma23prl,Yu25njp,Liu25NSR}.

The quantum metric integral (QMI), defined as $\mathcal{G}=\int_{\mathrm{BZ}}d^{2}\mathbf{k}\mathrm{Tr}[g(\mathbf{k})]$,
appears directly in a wide range of physical quantities. For instance,
in superconducting systems, the QMI contributes to the geometric part
of the superfluid weight and is essential for understanding flat-band
superconductivity \citep{Peotta15NC,Tovmasyan16prb,LiangL17prb,Herzog-Arbeitman22prl,ChenS24prl}.
In addition, it controls the spread of maximally localized Wannier
functions through the gauge-invariant spread function, thereby setting
a fundamental limit on electronic localization in real space \citep{Marzari97prb,Marziri12rmp}.
The QMI also governs the optical conductivity of band insulators \citep{Souza00prb,Yugo24prx,Ahn20prx},
determines the electron-phonon coupling constant in certain coupling
regimes \citep{YuJB24NP}, and contributes to the static dielectric
response \citep{Komissarov24NC}. These connections make the QMI a
broadly useful diagnostic of band geometry, with consequences that
extend beyond topological classifications.

A crucial result connecting quantum geometry and topology in 2D is
the inequality $\mathcal{G}\geq2\pi|C|$, where $C$ is the Chern
number; this bound follows from the pointwise relation $\mathrm{Tr}[g(\mathbf{k})]\geq|F_{xy}(\mathbf{k})|$
guaranteed by the positive semi-definiteness of the quantum geometric
tensor \citep{Roy14prb,Ozawa21prb}. This inequality establishes that
nonzero Chern number $C$ enforces a finite minimum amount of $\mathcal{G}$,
with profound implications: the superfluid weight, Wannier spread,
and optical conductivity are all bounded from below by the topological
invariant. Similar lower bounds have been established for time-reversal
symmetric systems, where the $\mathbb{Z}_{2}$ invariant \citep{YuJB25prl,Jankowski25prr}
and the Euler number $e_{2}$ \citep{YuJB23prb,Kwon24prb,Jankowski25prb}
impose analogous constraints on the $\mathcal{G}$. However, this
logic appears to fail for the large class of systems that are topologically
trivial in the conventional 2D sense \textemdash{} those with vanishing
Chern number $C=0$ or even identically vanishing Berry curvature
$F_{xy}(\mathbf{k})=0$ throughout the Brillouin zone. For such systems,
the standard inequality reduces to $\mathcal{G}\geq0$, imposing no
geometric constraints. This raises the central question of the present
work: can we generalize the relation between quantum geometry and
topology such that even the QMI $\mathcal{G}$ in trivial 2D systems
can still acquire a nonzero lower bound?

In this work, we answer this question affirmatively by demonstrating
that a tight, nonzero lower bound on the QMI can emerge in such trivial
2D systems from topological obstructions in lower dimensions. The
key insight is a dimensional-reduction technique: by decomposing the
2D QMI $\mathcal{G}$ into two nested 1D integrals along different
momenta $k_{x}$ and $k_{y}$, the dimension-reducted quantum metric
components $\mathcal{G}_{x}(k_{y})$ and $\mathcal{G}_{y}(k_{x})$
are independently bounded by 1D topological obstructions, e.g., the
winding numbers protected by chiral symmetry, or quantized Wannier
bands. This dimension-reduction strategy mirrors the philosophy of
nested Wilson loops used to characterize higher-order topological
phases, and reveals that the quantum geometry of a 2D system retains
a nontrivial imprint of its lower-dimensional topological structure.
We explicitly demonstrate our theory in three different models. We
first establish this mechanism in weak topological phases protected
by chiral symmetry. In the proposed tilted 2D Su-Schrieffer-Heeger
(SSH) model with vanishing Berry curvature, the QMI $\mathcal{G}$
acquires a nonzero lower bound from anisotropic 1D topology. It takes
the form $\mathcal{G}\geq4\pi^{2}(\nu_{x}^{2}+\nu_{y}^{2})$ in terms
of quantized Wannier centers $\nu_{x}$ and $\nu_{y}$. We then show
that the same logic in an anisotropic Wilson-Dirac model, where chiral
symmetry is present only on selected momentum slices. In this case,
the lower bound is generalized to $\mathcal{G}\geq2\pi^{2}(P_{x}+P_{y})$
in terms of the defined polarization weight $P_{x}$ and $P_{y}$.
Finally, we show that this framework extends naturally to higher-order
topological insulators (HOTIs) with zero Chern number or vanishing
Berry curvature. We introduce a Wannier-band quantum metric $\mathcal{G}^{w}$
defined on the Wannier-band basis in the Benalcazar-Bernevig-Hughes
(BBH) model, rather than in the original Bloch band basis. This Wannier-band
QMI reflects the topological phase diagram faithfully, diverging at
phase boundaries where the Wannier-band gap closes, and satisfies
the lower bound $\mathcal{G}^{w}\geq4\pi^{2}q_{xy}$, connecting directly
to the quadrupole moment $q_{xy}$. Our results generalize the relations
between quantum geometry and topology, providing quantitative and
physically meaningful bounds on QMI in regimes where conventional
topological arguments give no useful information.

The rest of the article is organized as follows. In Sec. II, we present
the dimension-reduction method to calculate the QMI following a nested-loop
approach and show the possibility to get nonzero lower bound. In Sec.
III, we present the titled 2D SSH model and show the emergence of
lower bound of QMI by quantized Wannier centers even with zero Berry
curvature. In Sec. IV, we consider the anisotropic Wilson-Dirac model
and generalize the lower-bound results with finite electron polarizations.
In Sec. V, we define the Wannier-band quantum metric in HOTIs and
reveal its lower bound by the quadrupole moment in the BBH model.
In Sec. VI, we summarize our result and discuss the consequences of
the new lower bound of QMI.

\section{Dimension-reduction framework of quantum metric and lower-dimensional
topological constraints}

We first establish a general dimension-reduction framework that reveals
how the QMI of a 2D systems is constrained by lower-dimensional topological
obstructions. In terms of the projection operators, the quantum geometric
tensor is calculated as \citep{Resta11EJP,Graf21prb,Mitscherling25prb}

\begin{align*}
\chi_{\mu\nu}({\bf k}) & =\mathrm{Tr}[P({\bf k})\partial_{\mu}P({\bf k})\partial_{\nu}P({\bf k})],
\end{align*}
where $P({\bf k})=\sum_{n=1}^{N_{occ}}|\psi_{n}({\bf k})\rangle\langle\psi_{n}({\bf k})|$
is the projection operator onto the occupied $N_{\mathrm{occ}}$ bands,
$\mu,\nu=x,y$ are the spatial components, ${\bf k}=(k_{x},k_{y})$
is the Bloch momentum, and $\partial_{\mu}\equiv\partial/\partial_{k_{\mu}}$.
The quantum geometric tensor can be decomposed as

\begin{equation}
\chi_{\mu\nu}({\bf k})=g_{\mu\nu}({\bf k})-\frac{i}{2}F_{\mu\nu}({\bf k}),
\end{equation}
where the real symmetric part $g_{\mu\nu}({\bf k})$ accounts for
Fubini-Study metric measuring quantum distance between states under
infinitesimal change of ${\bf k}$, and the imaginary anti-symmetric
part $F_{\mu\nu}({\bf k})$ is Berry curvature. The quantum metric
and Berry curvature are closely related. The positive semidefiniteness
of the quantum geometric tensor gives the pointwise inequality $\mathrm{Tr}[g({\bf k})]\geq2\sqrt{\mathrm{det}g({\bf k})}\geq|F_{xy}({\bf k})|$
\citep{Roy14prb,Ozawa21prb}. After integration over the Brillouin
zone (BZ), one obtains the conventional topological lower bound

\begin{equation}
\mathcal{G}\equiv\int_{\mathrm{BZ}}d^{2}{\bf k}\mathrm{Tr}[g({\bf k})]\geq2\pi|C|,\label{eq:Inequality}
\end{equation}
where the Chern number $C=\frac{1}{2\pi}\int_{\mathrm{BZ}}d^{2}{\bf k}F_{xy}({\bf k})$.
We denote the quantum metric integral (QMI) as $\mathcal{G}$ in the
following. Crucially, this inequality indicates that QMI $\mathcal{G}$
has a lower bound determined by the first Chern number $C$. Importantly,
$\mathcal{G}$ is closely related to several physical quantities,
such as the Wannier function spread, superfluid weight, and optical
conductivity. Therefore, Eq\textcolor{black}{.\ \eqref{eq:Inequality}
suggests that the interested }physical quantities are also constrained
from below by nonzero Chern number in topologically nontrivial system.
However, the conventional inequality probes only the 2D Berry topology.
When $C=0$, it reduces to the trivial statement $\mathcal{G}\geq0$
\footnote{Note the the topological invariant such as the the $Z_{2}$ invariant
or Euler number $e_{2}$ in presence of time reversal symmetry can
also impose lower bound on $\mathcal{G}$.}. The same limitation is more pronounced when the Berry curvature
vanishes identically throughout the BZ. 

We now show that the QMI in such trivial 2D systems can still be lower
bounded by topological obstructions in lower-dimensional subspaces.
In general, the QMI can be decomposed as
\begin{alignat}{1}
\mathcal{G} & =\int_{\mathrm{BZ}}d^{2}{\bf k}\left[g_{xx}({\bf k})+g_{yy}({\bf k})\right]\nonumber \\
 & =\int_{0}^{2\pi}dk_{y}\mathcal{G}_{x}(k_{y})+\int_{0}^{2\pi}dk_{x}\mathcal{G}_{y}(k_{x}),\label{eq:QMI}
\end{alignat}
where the dimension-reducted QMIs are defined as

\begin{alignat}{1}
\mathcal{G}_{x}(k_{y}) & \equiv\int_{0}^{2\pi}dk_{x}g_{xx}({\bf k}),\nonumber \\
\mathcal{G}_{y}(k_{x}) & \equiv\int_{0}^{2\pi}dk_{y}g_{yy}({\bf k}).
\end{alignat}
In this way, the QMI in 2D is factorized to two nested 1D integrals
along $x$ and $y$ directions in a nested-loop way. The essential
point is that although $C=0$ or $F_{xy}({\bf k})=0$ in 2D, the defined
components $\mathcal{G}_{x}(k_{y})$ and $\mathcal{G}_{y}(k_{x})$
could obtain nonzero lower bounds due to the nontrivial topology in
the reduced 1D subspaces. By finding lower bounds of $\mathcal{G}_{x}(k_{y})$
and $\mathcal{G}_{y}(k_{x})$, we are able to get a new nontrivial
lower bound of the full 2D QMI $\mathcal{G}$. This dimension-reduction
method is conceptually analogous to the nested Wilson loop in characterizing
the higher-order topology \citep{Benalcazar17Science,BBH17prb,LiCA20prb}.
It also naturally captures the anisotropic topological properties
along different directions.

\subsection{New lower bound from quantized Wannier bands}

We now explicitly consider the dimension-reducted QMIs $\mathcal{G}_{x}(k_{y})=\int_{0}^{2\pi}dk_{x}g_{xx}({\bf k})$
and $\mathcal{G}_{y}(k_{x})\equiv\int_{0}^{2\pi}dk_{y}g_{yy}({\bf k})$,
which will be lower bounded by 1D topology. This 1D topology could
be characterized by winding number protected by chiral symmetry \footnote{It is also possible for $Z_{2}$ invariant protected by the particle-hole
symmetry}. For simplicity, we first consider the component $\mathcal{G}_{x}=\int dk_{x}g_{xx}(k_{x})$
by treating $k_{y}$ as an independent parameter. To be concrete,
we consider a two-band insulator with chiral symmetry. The general
Hamiltonian is

\begin{equation}
H_{A}(k)=d_{x}(k)\sigma_{x}+d_{y}(k)\sigma_{y}=\left(\begin{array}{cc}
0 & de^{-i\phi(k)}\\
de^{i\phi(k)} & 0
\end{array}\right),
\end{equation}
where $d=\sqrt{d_{x}^{2}+d_{y}^{2}}$ and $\phi(k)=\arctan(d_{y}/d_{x})$.
For the occupied band, the quantum metric is 

\begin{alignat}{1}
g_{xx}(k) & =\frac{1}{2}\mathrm{Tr}[\partial_{k}P\partial_{k}P]=\frac{1}{4}(\partial_{k}\phi)^{2}.
\end{alignat}
The corresponding 1D QMI is $\int_{0}^{2\pi}dkg_{xx}(k)=\frac{1}{4}\int_{0}^{2\pi}dk(\partial_{k}\phi)^{2}.$
Using the Cauchy-Schwartz inequality $\int dk(f)^{2}\int1dk\geq(\int dkf)^{2}$
, we obtain
\begin{alignat}{1}
\mathcal{G}_{x}=\int_{0}^{2\pi}dkg_{xx}(k) & \geq\frac{\pi}{2}\left(\frac{1}{2\pi}\int_{0}^{2\pi}dk\partial_{k}\phi\right){}^{2}=\frac{\pi}{2}w^{2},
\end{alignat}
where $w$ is the winding number $w=\frac{1}{2\pi}\int_{0}^{2\pi}dk\partial_{k}\phi$.
It is consistent with the multi-band result \citep{Tovmasyan16prb},
as also discussed in Appendix E. Restoring the transverse momentum
dependence gives $\mathcal{G}_{x}(k_{y})\geq\frac{\pi}{2}w^{2}(k_{y})$.
Considering the relation from Wannier bands and winding number, this
bound can equivalently be written as
\begin{equation}
\mathcal{G}_{x}(k_{y})\geq2\pi\nu_{x}^{2}(k_{y}),
\end{equation}
where $\nu_{x}(k_{y})=\frac{w(k_{y})}{2}$ mod 1 is the Wannier bands
for the occupied state that can be obtained from the Wilson loop (see
Appendix C). Similar result can be obtained along the other direction
$\mathcal{G}_{y}(k_{x})\geq2\pi\nu_{y}^{2}(k_{x})$.

For certain band insulators, the Wannier band could be flat $\nu_{x}(k_{y})=\nu_{x}$
with quantized value $\nu_{x}=0,\frac{1}{2}$, as shown in Fig.\ \ref{fig:model I}(c)
and \ref{fig:model I}(d). The absence of Wannier-band winding is
consistent with a zero Chern number, but the nonzero Wannier centers
still represent 1D topological obstructions. Substituting the reduced
bounds into the 2D QMI gives

\begin{alignat}{1}
\mathcal{G} & =\int_{0}^{2\pi}dk_{y}\mathcal{G}_{x}(k_{y})+\int_{0}^{2\pi}dk_{x}\mathcal{G}_{y}(k_{x})\nonumber \\
 & \geq4\pi^{2}(\nu_{x}^{2}+\nu_{y}^{2}).\label{eq:WindingBound}
\end{alignat}
This result explicitly shows that a 2D system with $C=0$, or even
vanishing Berry curvature, can still possess a finite topological
lower bound on its quantum geometry. If both directions are topologically
nontrivial, $\nu_{x}=\nu_{y}=\frac{1}{2}$, then $\mathcal{G}\geq2\pi^{2}$.
This new lower bound is substantially tighter than the $2\pi|C|$
even if $C=\pm1$. If only one direction is topologically nontrivial,
namely $\nu_{x}=\frac{1}{2}$ and $\nu_{y}=0$, or vice versa, we
have $\mathcal{G}\geq\pi^{2}$. Both cases give rise to a tighter
lower bound than the original trivial zero limit $\mathcal{G}\geq0$.

\subsection{New lower bound from finite polarizations }

We next relax the requirement of chiral symmetry. Consider a general
two-band Hamiltonian in 1D $H_{B}(k)={\bf d}(k)\cdot{\bf \sigma}$
where ${\bf d}(k)=(d_{x},d_{y},d_{z})$ and the Pauli matrix vector
${\bf \sigma}=(\sigma_{x},\sigma_{y},\sigma_{z})$. As the momentum
$k$ goes around the BZ, the $\hat{d}$ vector $\hat{{\bf d}}={\bf d}(k)/|{\bf d}(k)|$
forms a closed loop on the sphere $S^{2}$ with a unit radius $R=1$.
Thus the infinitesimal distance on the loop is $dl=|\partial_{k}\hat{{\bf d}}|dk$
\citep{LiuT26prl}, leading to the total length 
\begin{equation}
L=\oint dl=\int_{0}^{2\pi}|\partial_{k}\hat{{\bf d}}|dk.
\end{equation}
The quantum metric for this model is $g_{xx}=\frac{1}{4}(\partial_{k}\hat{{\bf d}})^{2}$
\citep{Graf21prb}. The corresponding quantum distance is given by
\begin{alignat}{1}
\ell\equiv\int_{0}^{2\pi}\sqrt{g_{xx}(k)}dk & =\int_{0}^{2\pi}\frac{1}{2}|\partial_{k}\hat{{\bf d}}|dk,
\end{alignat}
which is one half of $L$, i.e., $L=2\ell$. Applying the Cauchy-Schwartz
inequality leads to
\begin{alignat}{1}
\mathcal{G}_{x}=\int_{0}^{2\pi}g_{xx}dk & \geq\frac{\ell^{2}}{2\pi}.
\end{alignat}
The closed curve on the Bloch sphere encloses a solid angle $\Omega$.
The spherical isoperimeteric inequality $L^{2}\geq\Omega(4\pi-\Omega)$
\citep{ZhangF26prl,Osserman78math}, together with $L=2\ell$, leads
to $\ell^{2}\geq\pi\Omega-\frac{\Omega^{2}}{4}.$ Considering the
relation $\Omega=2\gamma$ with $\gamma$ the Berry phase enclosed
by the loop \citep{XiaoD10rmp}, one obtains a general relation between
the quantum distance and Berry phase $\ell^{2}\geq\gamma(2\pi-\gamma)$,
where the Berry phase takes value in the region $[0,2\pi]$. Note
that the Wannier center and the Berry phase can be further related
as $\nu=\frac{\gamma}{2\pi}$, yielding 

\begin{alignat}{1}
\ell^{2} & \geq4\pi^{2}\nu(1-\nu).
\end{alignat}
Employing sequence of inequality, we arrive at
\begin{align}
2\pi\int_{0}^{2\pi}g_{xx}dk & \geq\ell^{2}\geq4\pi^{2}\nu(1-\nu).
\end{align}
Combining these two inequalities yields the interested formula 
\begin{align}
\mathcal{G}_{x} & \geq2\pi\nu(1-\nu).
\end{align}
It indicates that the 1D QMI can be bounded by nonzero Wannier center
in a weak condition than that requiring chiral symmetry\footnote{When $\nu$ is quantized at $\nu=0,\frac{1}{2}$, this formula reduced
to the previous strong result $\mathcal{G}_{x}\geq\frac{\pi}{2}\nu^{2}$}. The essential point is that, even when the Wannier bands are not
globally quantized, crystalline or internal symmetries can still pin
them to $\nu=\frac{1}{2}$ at specific momenta, as shown in Fig.\ \ref{fig:model_II}
(b). Such symmetry-protected constraints prevent the Wannier bands
from being continuously deformed to the atomic limit and ensure a
finite integrated geometric weight. Therefore, integrating the reduced
bound over the transverse momentum will give a finite and nonzero
value $\int_{0}^{2\pi}dk_{y}2\pi\nu_{x}(k_{y})(1-\nu_{x}(k_{y}))\equiv2\pi^{2}P_{x}>0$
with $P_{x}$ the defined polarization weight. Similar consideration
applies to the other direction $\int_{0}^{2\pi}dk_{x}2\pi\nu_{y}(k_{x})(1-\nu_{y}(k_{x}))=2\pi^{2}P_{y}>0$.
Therefore, the total 2D QMI is then bounded as

\begin{alignat}{1}
\mathcal{G} & =\int_{0}^{2\pi}dk_{y}\mathcal{G}_{x}(k_{y})+\int_{0}^{2\pi}dk_{x}\mathcal{G}_{y}(k_{x})\nonumber \\
 & \geq2\pi^{2}(P_{x}+P_{y})\gg0.\label{eq:Polarization_bound}
\end{alignat}
Although $P_{x}$ and $P_{y}$ are not quantized in general, they
remain finite when the Wannier bands are constrained by lower-dimensional
topological obstructions. It indicates that the symmetry-constrained
1D polarization can still impose a meaningful lower bound on the 2D
quantum geometry, even in the absence of chiral symmetry.

\section{Dimension-reducted quantum metric in model I: a tilted 2D SSH model}

\begin{figure}
\includegraphics[width=1\linewidth]{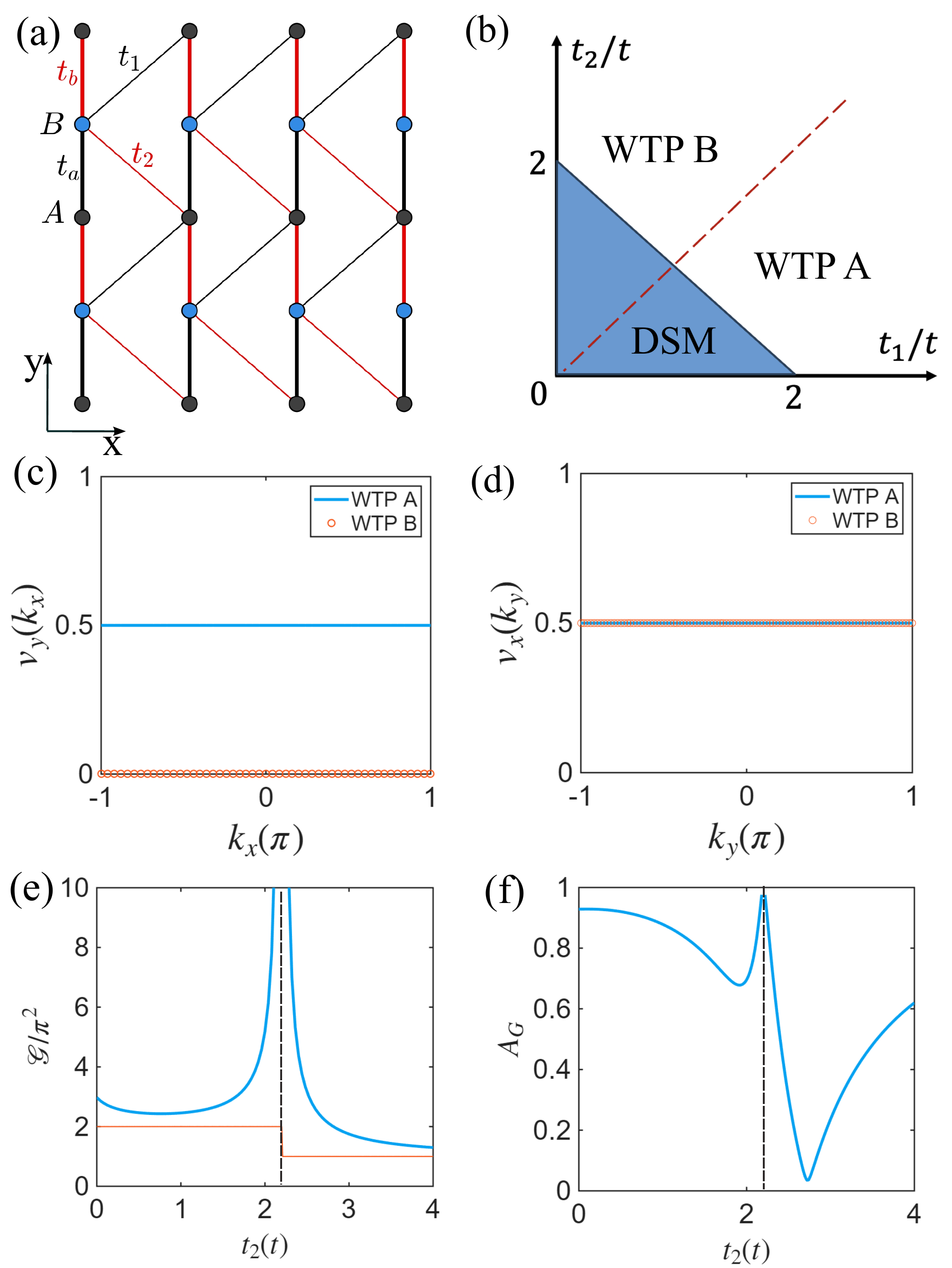}

\caption{(a) Sketch of the lattice for the tilted 2D SSH model. Dimerized hoppings
are labeled by $t_{1}-t_{a}$ and $t_{2}-t_{b}$ (thick vs thin lines)
along different directions (red vs blue color). For our purpose, we
consider the simple case with $t_{a}=t_{b}=t$. (b) Phase diagram
of the tilted 2D SSH model lattice in the parameter space $(t_{1},t_{2})$.
The blue region is Dirac semimetal (DSM), and the remaining insulating
phases are divided to two weak topological phases WTP A and WTP B
by the nodal-line semimetal phase (dashed red line). (c) The Wannier
band $\nu_{y}$ as a function of $k_{x}$ for WTP A and WTP B. (d)
The Wannier band $\nu_{x}$ as a function of $k_{y}$ similar to (c).
(e) The QMI $\mathcal{G}$ as a function of $t_{2}$ at fixed parameter
$t_{1}=2.2t$. The solid red line denotes the new lower bound of the
QMI and the dashed black line indicates the phase boundary. (f) The
quantum metric anisotropy $A_{G}$ as a function of $t_{2}$ at $t_{1}=2.2t$.
\label{fig:model I}}
\end{figure}

We now illustrate the dimension-reduction framework in a tilted 2D
SSH model. This model provides a particularly transparent example
because its Berry curvature vanishes identically, while its 1D momentum-space
cuts can nevertheless possess nontrivial topology. The lattice structure
is shown in Fig.\ \ref{fig:model I}(a), where each unit cell contains
two different sites, $A$ and $B$. The dimerized hoppings along different
directions are illustrated by different colors. For simplicity, we
first consider the case with $t_{a}=t_{b}=t$ \footnote{For the case with $r=\frac{t_{a}}{t_{b}}\neq1$, it gives another
knob to tune the phase diagram}. The ratio $t_{a}/t_{b}$ provides an additional tuning parameter
but does not alter the central mechanism discussed below. The lattice
Hamiltonian is written as 

\begin{alignat}{1}
H_{1} & =\sum_{\mathbf{R}}(tC_{\mathbf{R},A}^{\dagger}C_{\mathbf{R},B}+tC_{\mathbf{R},B}^{\dagger}C_{\mathbf{R}+e_{y},A}+h.c.)\nonumber \\
 & +\sum_{\mathbf{R}}(t_{1}C_{\mathbf{R},B}^{\dagger}C_{\mathbf{R}+e_{x}+e_{y},A}+h.c.)\nonumber \\
 & +\sum_{\mathbf{R}}(t_{2}C_{\mathbf{R},B}^{\dagger}C_{\mathbf{R}+e_{x},A}+h.c.).
\end{alignat}
The operators $C_{\mathbf{R},\zeta}^{\dagger}$ ($C_{\mathbf{R},\zeta}$)
are creation (annihilation) operators at unit cell $\mathbf{R}$ and
sublattice $\zeta=A,B$. This model show highly anisotropic properties
by tuning the relative amplitude of $t_{1}$ and $t_{2}$. Note that
this model can be viewed as a deformation of the inclined 2D SSH model
\citep{LiCA22prr,LiCA22fp}. The Bloch Hamiltonian thus reads

\begin{alignat}{1}
H_{1}({\bf k}) & =d_{x}\sigma_{x}+d_{y}\sigma_{y},\nonumber \\
d_{x} & =[t\cos k_{y}+t+t_{2}\cos k_{x}+t_{1}\cos(k_{x}+k_{y})],\nonumber \\
d_{y} & =[t\sin k_{y}+t_{2}\sin k_{x}+t_{1}\sin(k_{x}+k_{y})].
\end{alignat}
Define $d_{x}+id_{y}=t(1+e^{ik_{y}})+e^{ik_{x}}(t_{2}+t_{1}e^{ik_{y}}).$
The Hamiltonian respects chiral symmetry $\sigma_{z}H_{1}({\bf k})\sigma_{z}^{-1}=-H_{1}({\bf k})$
with the chiral symmetry operator $\sigma_{z}$ as well as time-reversal
symmetry $\mathcal{T}H_{1}({\bf k})\mathcal{T}^{-1}=H_{1}(-{\bf k})$
where $\mathcal{T}=K$ is the complex conjugate operation. Therefore,
particle-hole symmetry is also preserved. The energy spectrum is $E_{\pm}({\bf k})=\pm\sqrt{d_{x}^{2}+d_{y}^{2}}=\pm|{\bf d}|=\pm d.$
The eigenstates take the form $|\psi_{\pm}\rangle=\frac{1}{\sqrt{2}}\left(\begin{array}{c}
1\\
\pm e^{i\phi}
\end{array}\right)$ with $e^{i\phi}=\frac{d_{x}+id_{y}}{d}$. Because the Bloch vector
lies entirely in the $d_{x}-d_{y}$ plane, the Berry curvature vanishes
wherever the spectrum is gapped.

The phase diagram is presented as in Fig.\ \ref{fig:model I}(b).
The system supports both insulating and semimetallic phases. We first
consider the insulating regime $|t_{1}+t_{2}|>2t$. The line $t_{1}=t_{2}$
separates two anisotropic weak topological phases (WTPs), denoted
WTP A and WTP B. Both WTP A and WTP B have zero Chern number, but
their dimension-reducted 1D Hamiltonians possess anisotropic topological
properties in different directions. The corresponding Wannier bands
along the two momentum directions are shown in Figs.\ \ref{fig:model I}(c)
and \ref{fig:model I}(d). In WTP A, the Wannier bands are pinned
at $\nu_{x}(k_{y})=\frac{1}{2}$ and $\nu_{y}(k_{x})=\frac{1}{2}$
for all momenta, indicating nontrivial 1D topological properties along
$x$ and $y$ directions although the 2D topology is trivial. The
corresponding energy spectra for a ribbon structure are presented
in Appendix A. The quantum metric for the model Hamiltonian $H_{1}({\bf k})$
can be obtained as $g_{\mu\nu}({\bf k})=\frac{1}{4}\frac{({\bf d}\times\partial_{\mu}{\bf d})\cdot({\bf d}\times\partial_{\nu}{\bf d})}{d^{4}}$
\citep{Graf21prb}. The resulting QMI $\mathcal{G}$ is calculated
as the integral in Eq\textcolor{black}{.\ \eqref{eq:QMI}.} To compare
the QMI with the dimension-reducted lower bound, we consider the parameter
cut at $t_{1}=2.2t$ in Fig.\ \ref{fig:model I}(b) . Along this
line cut, tuning $t_{2}$ drives the system from WTP A to WTP B through
the nodal-line critical point ($t_{2}=t_{1}$). The QMI is shown in
Fig.\ \ref{fig:model I}(e), together with the corresponding piecewise
lower bound denoted by red solid line. Clearly, in the WTP A phase,
the QMI $\mathcal{G}$ has a lower bound according to Eq\textcolor{black}{.\ \eqref{eq:WindingBound},
i.e., $\mathcal{G}\geq4\pi^{2}(\nu_{x}^{2}+\nu_{y}^{2})=2\pi^{2}$,
which is much stronger constraint than the previous zero value from
$C=0$. In WTP B phase, the Wannier bands demonstrate} $\nu_{x}(k_{y})=\frac{1}{2}$
while $\nu_{y}(k_{x})=0$ independent of momentum, indicating nontrivial
topological properties along $x$ but trivial along $y$ directions.
In this case, there is still a lower bound of the QMI as\textcolor{black}{{}
$\mathcal{G}\geq\pi^{2}$. }

\begin{figure}
\includegraphics[width=1\linewidth]{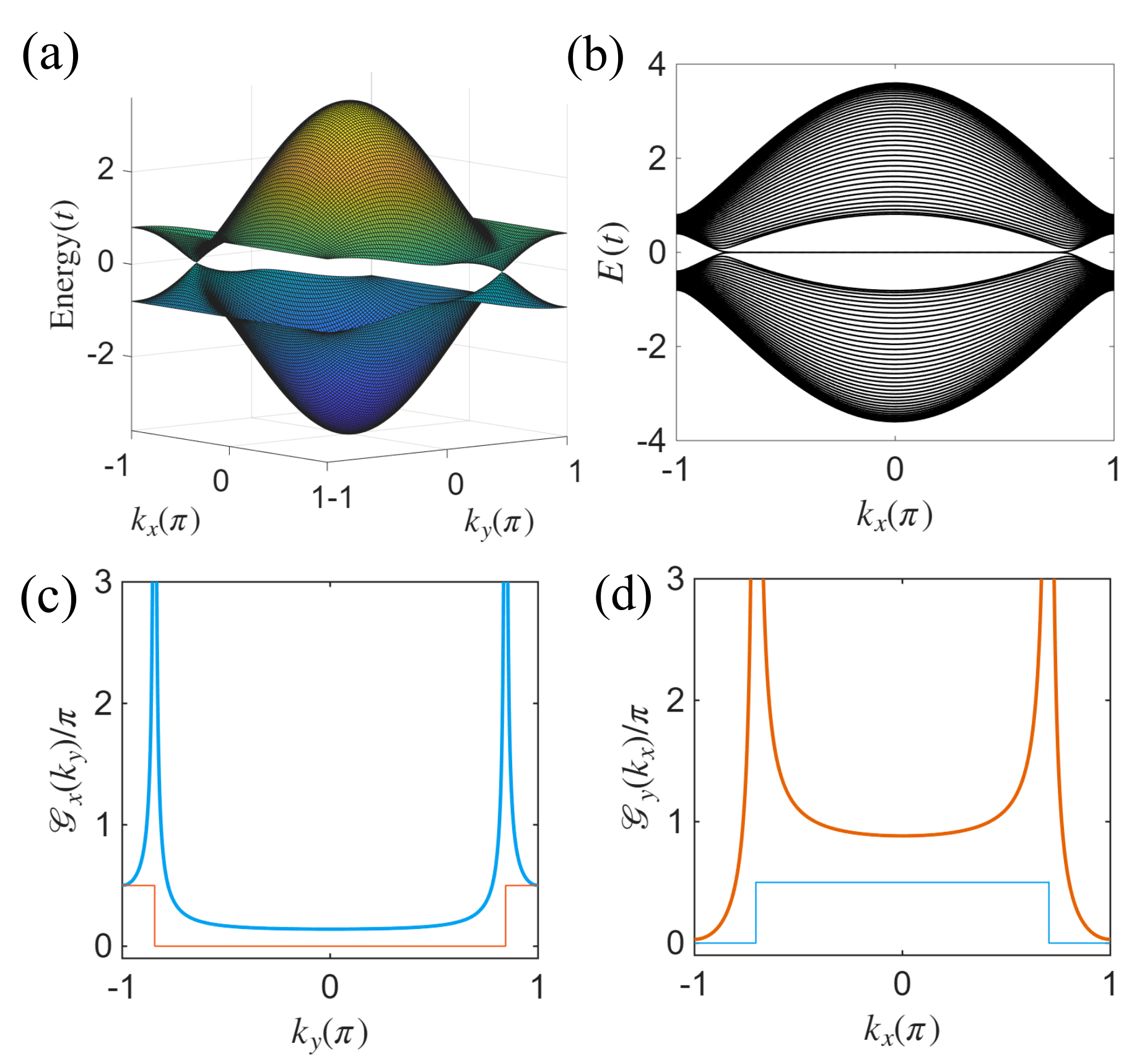}

\caption{(a) Band structure of the DSM in the tilted 2D SSH model. (b) Energy
spectrum on a ribbon along $x$ direction corresponding to (a). (c)
The dimension-reducted QMI $\mathcal{G}_{x}(k_{y})$ as a function
of $k_{y}$. The red line denotes the lower bound of QMI $\mathcal{G}_{x}(k_{y})$.
(d) The dimension-reducted QMI $\mathcal{G}_{y}(k_{x})$ as a function
of $k_{x}$. The blue line denotes the lower bound of QMI $\mathcal{G}_{y}(k_{x})$.
The parameters are taken at $t_{1}=1.2t,t_{2}=0.4t,$ and $t=1$.
\label{fig:model_I-Dirac}}
\end{figure}

\textcolor{black}{At the special line $t_{1}=t_{2}$, the system is
in a nodal-line phase. For simplicity, we set $t_{1}=t_{2}=\tau\geq0$
and $t=1$. The off-diagonal function factorizes as }$d_{x}+id_{y}=(t+\tau e^{ik_{x}})(1+e^{ik_{y}})$,
which becomes zero at $k_{y}=\pi$, giving rise to a nodal line for
the gapless phase. The projector for calculating quantum metric is
ill-defined on the nodal line since $d=0$ in this case. Introducing
a small mass term $\Delta\sigma_{z}$ regularizes the projector. As
derived in Appendix B, near $k_{y}=\pi$, the leading contribution
is $g_{yy}\simeq\frac{1}{4}\frac{\Delta^{2}v(k_{x})}{\left[\eta^{2}v(k_{x})+\Delta^{2}\right]^{2}}$,
where $\eta=k_{y}-\pi$. The singular region has width $\delta\eta\sim|\Delta|$,
while its peak height scales as $\Delta^{-2}$. Consequently, it gives
a divergent contribution to the metric integral $\mathcal{G}\sim\frac{1}{|\Delta|}$
as $\Delta\rightarrow0$ {[}see Appendix B{]}. This algebraic divergence
is stronger than the logarithmic divergence $\mathcal{G}\sim\log|\Delta|$
associated with an isolated Dirac point in 2D \citep{Yugo24prx}. 

The system exhibits highly anisotropic properties due to different
hopping amplitudes $t_{1}$ and $t_{2}$ along two directions. The
anisotropy will also be imprinted in the quantum metric. The quantum
metric $G_{\mu\nu}=\int d^{2}{\bf k}g_{\mu\nu}({\bf k})$ is positive
semidefinite and has two eigenvalues $\lambda_{\pm}$ with $\lambda_{+}\geq\lambda_{-}\geq0$.
These eigenvalues relates to the trace and determinant of $G$ as
$\mathrm{Tr}G=\mathcal{G}=\lambda_{+}+\lambda_{-}$, and $\mathrm{det}(G)=G_{xx}G_{yy}-G_{xy}^{2}=\lambda_{+}\lambda_{-}$.
We quantify the anisotropy by
\begin{equation}
A_{G}\equiv\frac{\lambda_{+}-\lambda_{-}}{\lambda_{+}+\lambda_{-}},
\end{equation}
which explicitly takes $A_{G}=\frac{\sqrt{(G_{xx}-G_{yy})^{2}+4G_{xy}^{2}}}{G_{xx}+G_{yy}}$.
By construction, $A_{G}$ lies in the interval $[0,1]$, where $A_{G}=0$
corresponds to isotropic quantum metric and $A_{G}=1$ corresponds
to highly anisotropic quantum metric. Figure \ref{fig:model I}(f)
shows $A_{G}$ along the same parameter cut used in Fig.\ \ref{fig:model I}(e).
It shows large values at $t_{2}=0$ and drop gradually as $t_{2}$
increases. While when $t_{2}$ approaches the nodal-line point $t_{1}^{c}=2.2t$,
the nodal-line singularity causes one metric eigenvalue to dominate,
which pushes $A_{G}$ back to $A_{G}\approx1$. Across the nodal line
phase, $A_{G}$ drops again while finally increases when $t_{2}$
is far away from $t_{1}^{c}$.

The Dirac semimetal phase is in the region $|t_{1}+t_{2}|<2t$ and
$t_{1}\neq t_{2}$. In this phase, two Dirac points locate at $K_{\pm}=\pm(k_{x}^{0},-k_{y}^{0})$,
where $\cos k_{x}^{0}=-\frac{t_{1}+t_{2}}{2t}$ and $\cos k_{y}^{0}=\frac{t_{1}^{2}+t_{2}^{2}-2t^{2}}{2(t^{2}-t_{1}t_{2})}$.
These Dirac points are not fixed at any high symmetry points but can
be continuously tuned by model parameters, as shown in Fig.\ \ref{fig:model_I-Dirac}(a).
There are flat edge bands connecting the two Dirac points when considering
the open boundary conditions {[}see Fig.\ \ref{fig:model_I-Dirac}(b){]}.
Because the bulk projector is singular at the Dirac points, the total
QMI diverges in the gapless limit. Instead, we show the dimension-reducted
quantities $\mathcal{G}_{x}(k_{y})$ and $\mathcal{G}_{y}(k_{x})$
in Fig.\ \ref{fig:model_I-Dirac}(c) and \ref{fig:model_I-Dirac}(d),
respectively. For the topologically nontrivial regions with $\nu_{x}(k_{y})=\frac{1}{2}$,
the reduced QMI satisfies $\mathcal{G}_{x}(k_{y})\geq\frac{\pi}{2}$.
Otherwise, the lower bound becomes zero. Similar result applies to
$\mathcal{G}_{y}(k_{x})$. Near the Dirac points, the dimension-reducted
quantum metric $\mathcal{G}_{x}(k_{y})$ and $\mathcal{G}_{y}(k_{x})$
both diverge.

\section{dimension-reducted quantum metric in model II: an anisotropic Wilson-Dirac
model}

\begin{figure}
\includegraphics[width=1\linewidth]{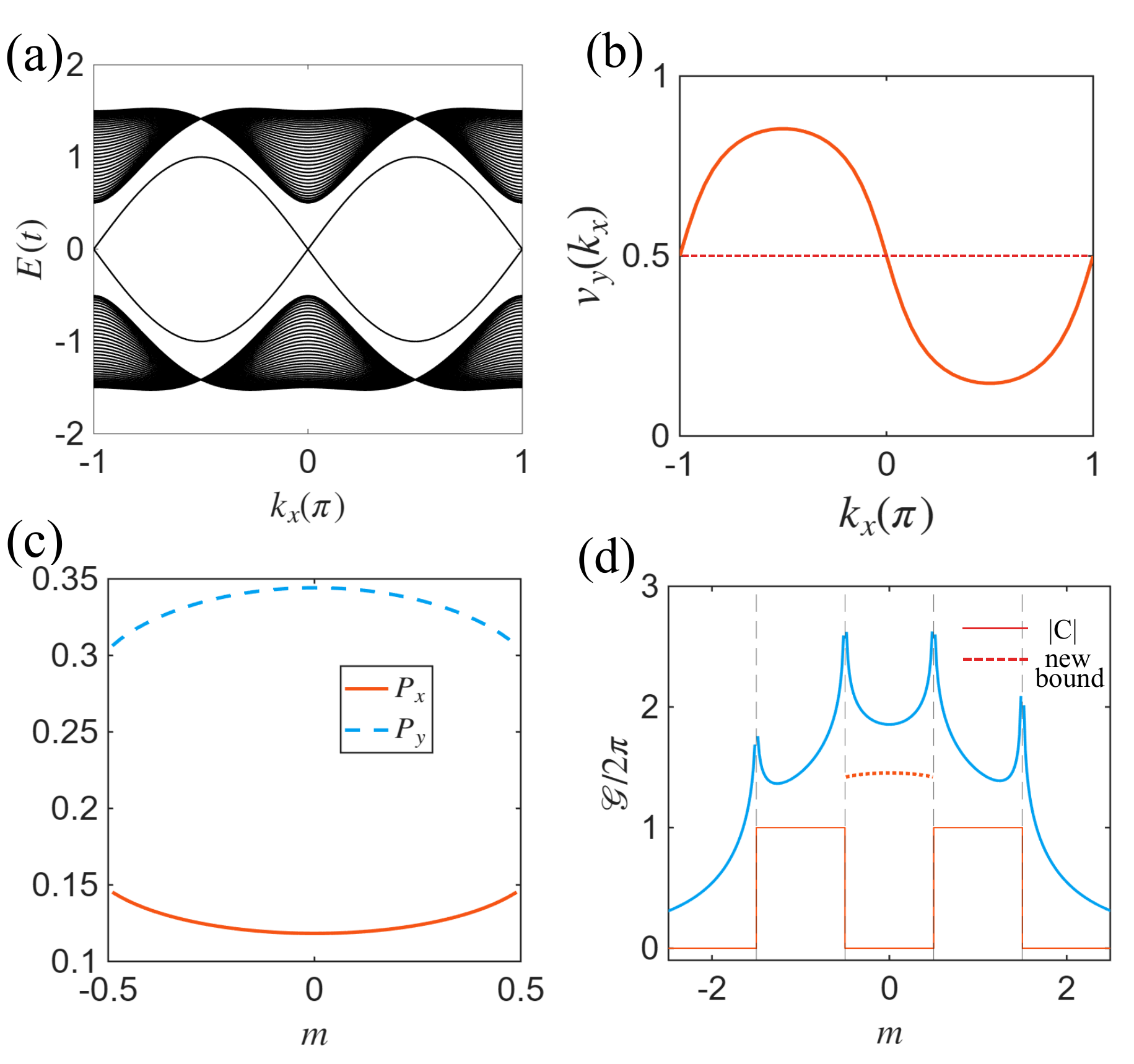}

\caption{(a) Energy spectrum of the anisotropic Wilson-Dirac model on a ribbon
along $x$ direction. There are a pair of edge states in the bulk
gap. Here $m=0$. (b) The Wannier bands $\nu_{y}$ as a function of
$k_{x}$. At the momentum $k_{x}=0,\pi$, the Wannier band is fixed
at $\nu_{y}=\frac{1}{2}.$ (c) The value of defined polarization weight
$P_{x}$ and $P_{y}$ for the QMI in Eq\textcolor{black}{.\ \eqref{eq:Polarization_bound}}.
(d) The QMI $\mathcal{G}$ and the corresponding Chern number $|C|$.
The dashed line indicates the new lower bound of QMI in the anisotropic
Wilson-Dirac model within $C=0$ region. The parameters are taken
at $b_{x}=0.5,b_{y}=1$, and $v=1$. \label{fig:model_II}}
\end{figure}

Next, we demonstrate that the dimension-reduction framework remains
applicable when chiral symmetry is absent globally while survives
only on isolated momentum-space cuts. To this end, we consider an
anisotropic Wilson-Dirac model with the Bloch Hamiltonian

\begin{alignat}{1}
H_{2}({\bf k}) & ={\bf d}({\bf k})\cdot\bm{\sigma},\nonumber \\
{\bf d}({\bf k}) & \equiv(v\sin k_{x},v\sin k_{y},m+\sum_{i=x,y}b_{i}\cos k_{i}).\label{eq:Model II}
\end{alignat}
where $m$ is the mass term, $b_{x,y}$, and $v$ are model parameters.
The anisotropy of this model arises from the choice $b_{x}\neq b_{y}$
\citep{LiCA25prb}. It is a generalization of the Chern insulator
model \citep{Qi06prb,Bernevig06Scien}. The Chern number can be calculated
as $C=\int_{BZ}\frac{d^{2}{\bf k}}{4\pi|{\bf d}({\bf k})|^{3}}{\bf d}({\bf k})\cdot\partial_{x}{\bf d}({\bf k})\times\partial_{y}{\bf d}({\bf k})$.
For simplicity, we define $b_{\pm}\equiv|b_{x}\pm b_{y}|$ and assume
$b_{+}>b_{-}$. The Chern number takes $C=+1$ for $-b_{+}<m<-b_{-}$,
$C=-1$ for $b_{-}<m<b_{+}$, and $C=0$ otherwise. Notably, an additional
phase with $C=0$ emerges for $|m|<b_{-}$ if $b_{x}\neq b_{y}$.
In this phase, a pair of edge states appears along one boundary but
not the other boundary {[}see Fig.\ \ref{fig:model_II}(a){]}. It
is characteristic of a WTP, where the topological protection and boundary
states depend on specific crystal directions \citep{Yoshimura14prb,FuL07prb,Hughes11prb}. 

The quantum metric for this model can be obtained as $\ensuremath{g_{\mu\nu}=\frac{1}{4}\left[\frac{\partial_{\mu}\mathbf{d}\cdot\partial_{\nu}\mathbf{d}}{|\mathbf{d}|^{2}}-\frac{\left(\mathbf{d}\cdot\partial_{\mu}\mathbf{d}\right)\left(\mathbf{d}\cdot\partial_{\nu}\mathbf{d}\right)}{|\mathbf{d}|^{4}}\right]}$\citep{Graf21prb}.
The QMI $\mathcal{G}$ in different phase regions is plotted in Fig.\ \ref{fig:model_II}(d).
As expected, the conventional inequality $\mathcal{G}\geq2\pi|C|$
is satisfied. In the WTP region $|m|<b_{-}$, however, it reduces
to the trivial bound $\mathcal{G}\geq0$, whereas the calculated QMI
remains substantially finite. We now show that this finite geometric
weight is captured by the momentum-dependent Wannier bands of the
dimension-reducted Hamiltonians. In the WTP phase, the corresponding
Wannier band along $y$ direction is presented in Fig.\ \ref{fig:model_II}(b).
It goes through a trivial path with zero winding number, consistent
with the vanishing Chern number. However, at the momentum $k_{x}^{*}=0,\pi$,
the Wannier center is fixed at $\nu_{y}(k_{x}^{*})=\frac{1}{2}$.
This is due to the chiral symmetry of the dimension-reducted Hamiltonian
$H(k_{x}^{*},k_{y})=(m\pm b_{x}+b_{y}\cos k_{y})\sigma_{x}+v\sin k_{y}\sigma_{y}$
\citep{LiCA25prb}. The Wannier band $\nu_{y}(k_{x})$ satisfies $\nu_{y}(k_{x})+\nu_{y}(-k_{x})=0\ \mathrm{mod}\ 1$
{[}see Appendix D{]}. From the above results, we have $\mathcal{G}_{y}(k_{x})\geq2\pi\nu_{y}(k_{x})[1-\nu_{y}(k_{x})]$.
Thus the integral satisfies

\begin{align}
\int_{-\pi}^{\pi}dk_{x}\mathcal{G}_{y}(k_{x}) & \geq4\pi\int_{0}^{\pi}dk_{x}[\nu_{y}(k_{x})-\nu_{y}^{2}(k_{x})]\nonumber \\
 & =2\pi^{2}\left(\frac{1}{2}-\frac{2}{\pi}\int_{0}^{\pi}\delta^{2}(k_{x})dk_{x}\right)=2\pi^{2}P_{y},
\end{align}
where $\delta(k_{x})\equiv\frac{1}{2}-\nu_{y}(k_{x})$ is the deviation
of the Wannier bands from the quantized value $\nu=\frac{1}{2}$.
The result goes back to the previous one $\int_{-\pi}^{\pi}dk_{x}\mathcal{G}_{y}(k_{x})\geq\pi^{2}$
for $\delta(k_{x})=0$. For Wannier band along the $y$ direction,
the Wannier center is fixed at $\nu_{x}(k_{y}^{*})=0$ at the momenta
$k_{y}^{*}=0,\pi$. Thus the integral gives $\int_{-\pi}^{\pi}dk_{y}\mathcal{G}_{x}(k_{y})\geq4\pi\int_{0}^{\pi}dk_{y}[\nu_{x}(k_{y})-\nu_{x}^{2}(k_{y})]=2\pi^{2}P_{x}$.
Therefore, the QMI $\mathcal{G}$ obeys a new lower bound

\begin{equation}
\mathcal{G}\geq2\pi^{2}(P_{x}+P_{y})\gg0.\label{eq:QMI_polarization}
\end{equation}
Although $P_{x}$ and $P_{y}$ are not quantized in general, they
cannot be suppressed to zero due to the anisotropic topological obstruction
{[}see Fig.\ \ref{fig:model_II}(c){]}. The numerical QMI and the
corresponding dimension-reducted lower bound are compared in Fig.\ \ref{fig:model_II}(d).
In the WTP, the new bound is finite and substantially tighter than
the conventional zero Chern number bound. 

\section{Wannier-band quantum metric in higher-order topological insulators:
The BBH model}

The connection between topology and quantum metric has been substantially
investigated in the conventional first-order topological phases while
it is not explored yet in the higher-order topological phases. With
the dimension-reduction framework, this connection could be generalized
to higher-order topological insulators (HOTIs) with zero Chern number
or vanishing Berry curvature as we show below. In HOTIs, the relevant
topological obstruction is encoded in nested Wannier bands. The prototypical
model of HOTIs is the BBH model \citep{Benalcazar17Science,BBH17prb}

\begin{alignat}{1}
H_{3}({\bf k}) & =t\sin k_{y}\gamma_{1}+(t_{y}+t\cos k_{y})\gamma_{2}\nonumber \\
 & +t\sin k_{x}\gamma_{3}+(t_{x}+t\cos k_{x})\gamma_{4},
\end{alignat}
where the $\gamma$ matrices are defined as $\gamma_{j}=-\tau_{2}\sigma_{j}$
with $j=1,2,3$ and $\gamma_{4}=\tau_{1}\sigma_{0}$ with $\tau$
and $\sigma$ both being Pauli matrices. The model respects chiral
symmetry $\gamma_{5}H_{3}({\bf k})\gamma_{5}^{-1}=-H_{3}({\bf k})$,
where the chiral symmetry operator $\gamma_{5}\equiv-\gamma_{1}\gamma_{2}\gamma_{3}\gamma_{4}=\tau_{3}\sigma_{0}$
\citep{LiCA20prl}. With the help of chiral symmetry, energy spectra
can be obtained as $E^{\pm}({\bf k})=\pm\sqrt{\epsilon_{x}^{2}(k_{x})+\epsilon_{y}^{2}(k_{y})}$
where $\epsilon_{y}^{2}(k_{y})\equiv t_{y}^{2}+2t_{y}t\cos k_{y}+t^{2}$
and similarly for $\epsilon_{x}(k_{x})$. Each band is twofold degenerate.
The system is gapped and the Chern number is zero. Its topology is
instead characterized by the quadrupole moment $q_{xy}$. In the nontrivial
phase, defined by $|t_{x}/t|<1$ and $|t_{y}/t|<1$, the quadrupole
moment takes $q_{xy}=\frac{1}{2}$. Otherwise, $q_{xy}=0$ in the
trivial regions.

\begin{figure}
\includegraphics[width=1\linewidth]{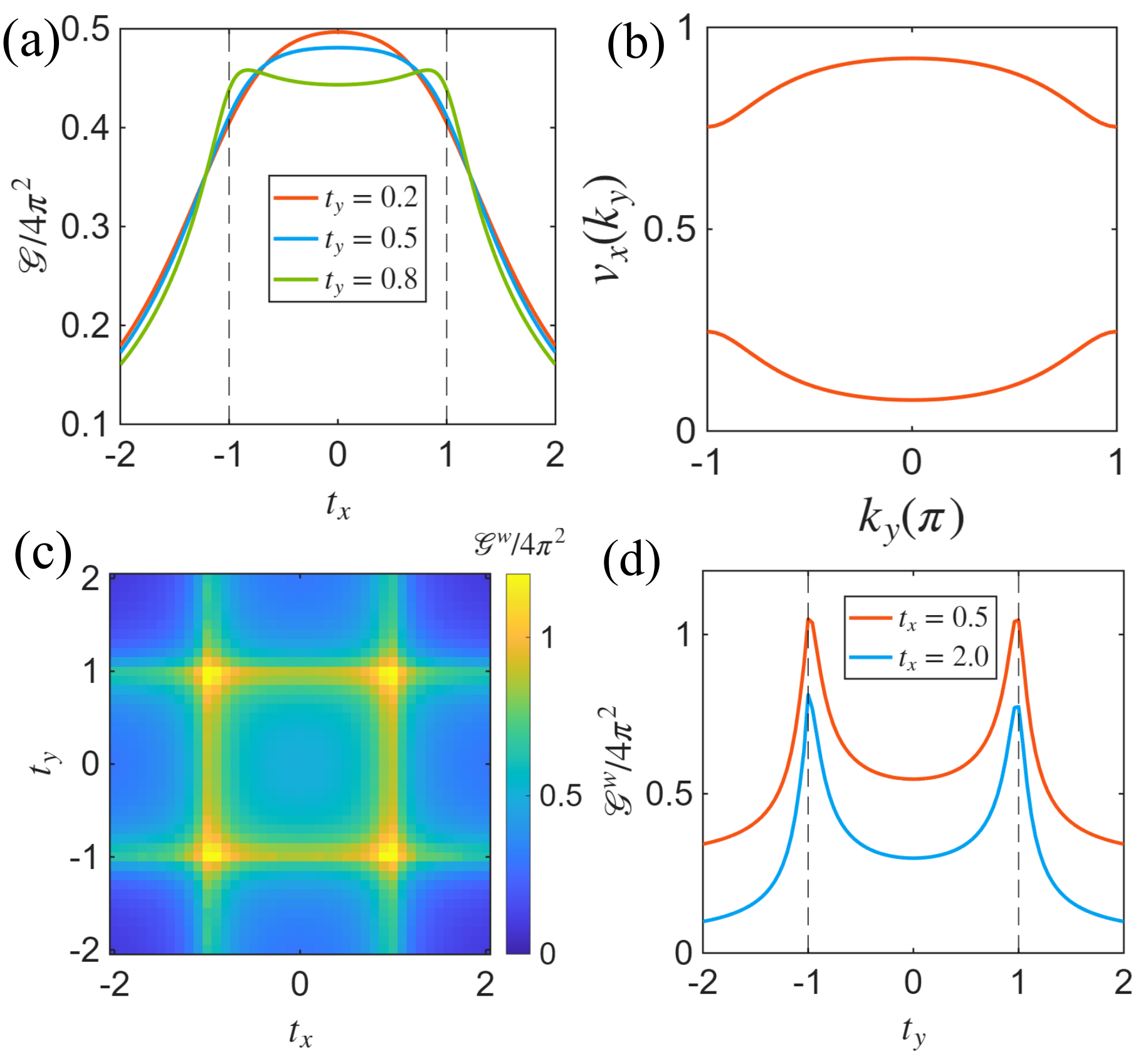}

\caption{(a) The Bloch-basis QMI $\mathcal{G}$ as a function of $t_{x}$ for
different $t_{y}$ in the BBH model. The dashed lines indicate the
topological phase boundaries. (b) Wannier bands of the BBH model at
$t_{x}=t_{y}=0.5t$. (c) The phase diagram of the Wannier-band basis
QMI $\mathcal{G}^{w}$ in the parameter space $(t_{x},t_{y})$. In
the topological regions, QMI $\mathcal{G}^{w}$ is lower bounded by
non-zero quadrupole moment. It becomes divergent at the phase boundaries.
(d) Wannier-basis QMI $\mathcal{G}^{w}$ as a function of $t_{y}$
at fixed $t_{x}=0.5t$ and $t_{x}=2.0t$. We set $t=1$ for all plots.
\label{fig:model_HOTI}}
\end{figure}

The quantum metric for occupied states can be obtained from $g_{\mu\nu}({\bf k})=\frac{1}{2}\mathrm{Tr}[\partial_{\mu}P({\bf k})\partial_{\nu}P({\bf k})]$,
where $P({\bf k})=\sum_{n=1}^{2}|\psi_{n}({\bf k})\rangle\langle\psi_{n}({\bf k})|$
is the projection operator at half-filling. We find that $\mathcal{G}$
takes much larger values in the topological nontrivial regions, while
it does not provide a direct connection to $q_{xy}$. Indeed, the
dimension-reducted QMI for multi-band systems can be given as

\begin{equation}
\mathcal{G}_{x}(k_{y})\geq\frac{\pi}{2N_{\mathrm{occ}}}w_{x}^{2}(k_{y}),
\end{equation}
where $w_{x}(k_{y})$ is the winding number of occupied bands and
$N_{\mathrm{occ}}=2$ {[}see Appendix E{]}. In the BBH model, this
winding number $w_{x}(k_{y})=0$ due to the requirement of well-defined
quadrupole moment. Similarly, $\mathcal{G}_{y}(k_{x})\geq\frac{\pi}{2N_{\mathrm{occ}}}w_{y}^{2}(k_{x})$
with $w_{y}(k_{x})=0$. Therefore, the total Bloch-basis QMI $\mathcal{G}$
retains only the trivial lower bound, i.e. $\mathcal{G}\geq0$, consistent
with the zero Chern number and vanishing Berry curvature in this model.
Besides, the Bloch-basis QMI does not sharply resolve the boundaries
between nontrivial and trivial HOTI phase. As shown in Fig.\ \ref{fig:model_HOTI}(a),
the QMI $\mathcal{G}$ changes smoothly across the phase boundaries
and does not display divergent behavior. This is because the bulk
bands do not experience a gap close-reopen process during the phase
transition of HOTIs. In this case, there is no direct correspondence
between the Bloch-basis QMI $\mathcal{G}$ and the quadrupole moment
$q_{xy}$. It is reasonable since $\mathcal{G}$ is defined using
the Bloch wave function basis, while $q_{xy}$ is defined on the Wannier-band
basis.

To connect quantum geometry directly to the quadrupole moment $q_{xy}$,
we now construct a quantum metric in the Wannier-band basis. Take
the Wannier bands $\nu_{x}(k_{y})$ presented in Fig.\ \ref{fig:model_HOTI}(b)
as an example. Since these bands are gapped, we can use the states
below the Wannier band gap to construct a projection operator. The
Wilson loop determines the Wannier bands and the eigenstates through
$\mathcal{W}_{x,{\bf k}}|\nu_{x}^{\pm}(k_{y})\rangle=\exp[i2\pi\nu_{x}^{\pm}(k_{y})]|\nu_{x}^{\pm}(k_{y})\rangle$
\citep{Benalcazar17Science,BBH17prb}. Combining the Wannier-band
eigenvectors $|\nu_{x}^{-}(k_{y})\rangle$ with the occupied Bloch
wave functions, we construct the Wannier basis function $|w_{x,\mathbf{k}}^{\text{\ensuremath{\pm}}}\rangle=\sum_{n=1}^{N_{\mathrm{occ}}}\left|u_{\mathbf{k}}^{n}\right\rangle \left[\nu_{x,\mathbf{k}}^{\text{\ensuremath{\pm}}}\right]^{n}$.
The projector of the lower Wannier sector is then defined as $\mathcal{P}_{x}({\bf k})=|w_{x,\mathbf{k}}^{-}\rangle\langle w_{x,\mathbf{k}}^{-}|$.
A proper quantum metric in the Wannier-band basis can be defined as
\begin{equation}
g_{yy}^{w}({\bf k})=\frac{1}{2}\mathrm{Tr}[\partial_{y}\mathcal{P}_{x}({\bf k})\partial_{y}\mathcal{P}_{x}({\bf k})].
\end{equation}
Note that we have used a upper label $w$ to indicate the Wannier-band
basis and distinguish it from the Bloch basis. The corresponding dimension-reducted
QMI is then given by
\begin{equation}
\mathcal{G}_{y}^{w}(k_{x})=\int_{0}^{2\pi}g_{yy}^{w}({\bf k})dk_{y}.
\end{equation}
Similarly, we construct the Wannier-band basis $|w_{y,\mathbf{k}}^{-}\rangle=\sum_{n=1}^{N_{\mathrm{occ}}}|u_{\mathbf{k}}^{n}\rangle\left[\nu_{y,\mathbf{k}}^{-}\right]^{n}$
and the projector $\mathcal{P}_{y}({\bf k})=|w_{y,\mathbf{k}}^{-}\rangle\langle w_{y,\mathbf{k}}^{-}|$.
Then the other component of quantum metric is
\begin{equation}
g_{xx}^{w}({\bf k})=\frac{1}{2}\mathrm{Tr}[\partial_{x}\mathcal{P}_{y}({\bf k})\partial_{x}\mathcal{P}_{y}({\bf k})],
\end{equation}
which leads to dimension-reducted QMI 
\begin{alignat}{1}
\mathcal{G}_{x}^{w}(k_{y}) & =\int_{0}^{2\pi}g_{xx}^{w}({\bf k})dk_{x}.
\end{alignat}
Therefore, the total Wannier-band QMI reads
\begin{alignat}{1}
\mathcal{G}^{w} & =\int_{0}^{2\pi}dk_{y}\mathcal{G}_{x}^{w}(k_{y})+\int_{0}^{2\pi}dk_{x}\mathcal{G}_{y}^{w}(k_{x}).
\end{alignat}
Note that the Wannier-band quantum metric measures the quantum distance
of Wannier-band basis wave function under infinitesimal changes, while
its measurable consequences of QMI $\mathcal{G}^{w}$ needs further
investigations. The QMI $\mathcal{G}^{w}$ in the parameter space
$(t_{x},t_{y})$ is presented in Fig.\ \ref{fig:model_HOTI}(c).
It fits the topological phase region $|t_{x}|<1$ and $|t_{y}|<1$
with much large values. At the phase boundaries, it shows divergent
behavior, consistent with previous results. 

The Wannier-band QMI $\mathcal{G}^{w}$ is directly connected to the
higher-order topological invariant encoded by quadrupole moment $q_{xy}$.
Based on the Wannier-band basis function $|w_{x,\mathbf{k}}^{-}\rangle$,
we define the Wannier-band Berry connection as

\begin{equation}
\mathcal{A}_{y,\mathbf{k}}^{v_{x}}\equiv-i\left\langle w_{x,\mathbf{k}}^{-}\right|\partial_{y}\left|w_{x,\mathbf{k}}^{-}\right\rangle .
\end{equation}
Thus the dimension-reducted Wannier-band basis quantum metric follows
\begin{align}
\mathcal{G}_{x}^{w}(k_{y}) & =\int_{0}^{2\pi}g_{xx}^{w}({\bf k})dk_{x}\geq2\pi\left[\nu_{y}^{\nu_{x}}(k_{y})\right]^{2},
\end{align}
where 
\begin{align}
\nu_{y}^{\nu_{x}}(k_{y}) & =\frac{1}{2\pi}\int_{0}^{2\pi}\mathcal{A}_{y,\mathbf{k}}^{v_{x}}dk_{x}.
\end{align}
Then the integral specifically takes
\begin{align}
\int_{0}^{2\pi}\mathcal{G}_{x}^{w}(k_{y})dk_{y} & \geq2\pi\int_{0}^{2\pi}\left[\nu_{y}^{\nu_{x}}(k_{y})\right]^{2}dk_{y}\nonumber \\
 & \geq\left(\int_{0}^{2\pi}\nu_{y}^{\nu_{x}}(k_{y})dk_{y}\right)^{2}\nonumber \\
 & =\left[\int_{0}^{2\pi}\left(\frac{1}{2\pi}\int_{0}^{2\pi}\mathcal{A}_{y,\mathbf{k}}^{v_{x}}dk_{x}\right)dk_{y}\right]^{2}\nonumber \\
 & =(2\pi)^{2}\left(\frac{1}{(2\pi)^{2}}\int_{0}^{2\pi}\mathcal{A}_{y,\mathbf{k}}^{v_{x}}d^{2}{\bf k}\right)^{2}\nonumber \\
 & =4\pi^{2}\left[p_{y}^{\nu_{x}}\right]^{2},
\end{align}
where we have used the Cauchy-Schwartz inequality in the second line.
Similarly, the other component takes
\begin{align}
\int_{0}^{2\pi}\mathcal{G}_{y}^{w}(k_{x})dk_{x} & \geq2\pi\int_{0}^{2\pi}\left[\nu_{x}^{\nu_{y}}(k_{x})\right]^{2}dk_{x}\nonumber \\
 & =4\pi^{2}\left[p_{x}^{\nu_{y}}\right]^{2}.
\end{align}
Thus the total QMI reads
\begin{alignat}{1}
\mathcal{G}^{w} & =\int_{0}^{2\pi}dk_{y}\mathcal{G}_{x}^{w}(k_{y})+\int_{0}^{2\pi}dk_{x}\mathcal{G}_{y}^{w}(k_{x})\nonumber \\
 & \geq4\pi^{2}\times2p_{y}^{\nu_{x}}p_{x}^{\nu_{y}}=4\pi^{2}q_{xy},
\end{alignat}
where we have used the formula $q_{xy}=2p_{y}^{\nu_{x}}p_{x}^{\nu_{y}}$
\citep{Benalcazar17Science,BBH17prb}. The Wannier-band QMI $\mathcal{G}^{w}$
in the BBH model is thus indeed lower bounded by the quadrupole moment
$q_{xy}$. As shown in Fig.\ \ref{fig:model_HOTI}(c), the QMI satisfies
$\mathcal{G}^{w}\geq4\pi^{2}\times\frac{1}{2}$ in the topological
nontrivial region $|t_{x}|<1$ and $|t_{y}|<1$. It becomes more transparent
along the two representative lines cuts in Fig.\ \ref{fig:model_HOTI}(d).
For $t_{x}=0.5t,$ the $\mathcal{G}^{w}\geq4\pi^{2}\times\frac{1}{2}$
has a nonzero lower bound in the topological region $|t_{y}|<1$ while
it decays towards zero in the trivial regions as $|t_{y}|>1$. For
$t_{x}=2.0t$, there is no lower bound for $\mathcal{G}^{w}$ in the
whole parameter region. At the phase boundaries $|t_{y}|=1$, the
QMI $\mathcal{G}^{w}$ tends to be divergent. 

\section{Discussion and conclusions}

The new lower bound of QMI has direct impacts on some important physical
quantities of band insulators, such as the Wannier function spread
and the optical spectral weight. The Wannier function spread, $\Omega=\sum_{n}[\langle r^{2}\rangle_{n}-\langle r\rangle_{n}^{2}]$
with electron position operator $r$, denotes the standard derivation
of Wannier functions in real space. The gauge-invariant part of Wannier
function spread in 2D is given by \citep{Marzari97prb,Marziri12rmp}
\begin{alignat}{1}
\Omega_{\mathrm{I}} & =\frac{1}{(2\pi)^{2}}\int_{\mathrm{BZ}}d^{2}{\bf k}\mathrm{Tr}[g({\bf k})].
\end{alignat}
Henceforth, it gains a nonzero lower bound due to the new lower bound
of QMI as $\Omega_{\mathrm{I}}\geq\nu_{x}^{2}+\nu_{y}^{2}$ in model
I and $\Omega_{\mathrm{I}}\geq(P_{x}+P_{y})/2$ in model II following
Eq\textcolor{black}{.\ \eqref{eq:WindingBound}} and Eq\textcolor{black}{.\ \eqref{eq:Polarization_bound}},
respectively. These results are reasonable since the Wannier bands
determine the Wannier centers, the relative positions of electrons
with respect to the positive charge centers in the unit cell. The
Wannier function spread naturally obtain a minimum value obstructed
by nonzero $\nu_{x,y}$ and $P_{x,y}$, since the electrons cannot
be pushed to the atomic center due to the topological protection.

In addition, the optical conductivity of a band insulator is connected
to the QMI by the Souza-Wilkens-Martin formula \citep{Souza00prb}
\begin{alignat}{1}
\int_{0}^{\infty}\frac{d\omega}{\omega}\sum_{\mu}\mathrm{Re}\sigma_{\mu\mu}(\omega) & =\frac{e^{2}}{4\pi\hbar}\mathcal{G},
\end{alignat}
where $\sigma_{\mu\mu}(\omega)$ is the diagonal component of the
optical conductivity tensor. Therefore, the integral of optical conductivity
obtains a new lower bound $\int_{0}^{\infty}\frac{d\omega}{\omega}\sum_{\mu}\mathrm{Re}\sigma_{\mu\mu}(\omega)\geq\frac{e^{2}\pi}{\hbar}(\nu_{x}^{2}+\nu_{y}^{2})$
in model I and $\int_{0}^{\infty}\frac{d\omega}{\omega}\sum_{\mu}\mathrm{Re}\sigma_{\mu\mu}(\omega)\geq\frac{e^{2}\pi}{2\hbar}(P_{x}+P_{y})$
for model II, respectively, which are far different from the trivial
zero lower bound. For a band insulator with energy gap $\Delta$,
the optical conductivity is zero below the gap. Consequently, it follows
$\int_{0}^{\infty}\frac{d\omega}{\omega}\mathrm{Re}\sigma_{\mu\mu}(\omega)=\int_{\omega_{g}}^{\infty}\frac{d\omega}{\omega}\mathrm{Re}\sigma_{\mu\mu}(\omega)\leq\frac{\hbar}{\Delta}\int_{\omega_{g}}^{\infty}d\omega\mathrm{Re}\sigma_{\mu\mu}(\omega)=\frac{\hbar}{\Delta}\int_{0}^{\infty}d\omega\mathrm{Re}\sigma_{\mu\mu}(\omega)$
where $\hbar\omega_{g}=\Delta$. Using the optical sum rule $\int_{0}^{\infty}d\omega\mathrm{Re}\sigma_{\mu\mu}(\omega)=\frac{\pi}{2}\frac{\rho e^{2}}{m}$
with $\rho$ the electron density and $m$ the electron mass, it has
been shown that the energy gap has a upper bound from the QMI $\mathcal{G}$
as

\begin{align}
\Delta & \leq\frac{8\rho\pi^{2}\hbar^{2}}{m}\frac{1}{\mathcal{G}}.
\end{align}
Consequently, the new QMI lower bounds imply $\Delta\leq\frac{2\rho\hbar^{2}}{m}\frac{1}{\nu_{x}^{2}+\nu_{y}^{2}}$
for model I, and $\Delta\leq\frac{4\rho\hbar^{2}}{m}\frac{1}{P_{x}+P_{y}}$
for model II.

In summary, we have developed a dimension-reduction approach to evaluate
the QMI $\mathcal{G}$ in 2D systems with zero Chern number or vanishing
Berry curvature. Although these systems are topologically trivial
from the conventional viewpoint, we show that anisotropic 1D topology
protected by chiral symmetry can impose a nonzero lower bound on the
QMI $\mathcal{G}$. This lower bound arises from lower-dimensional
topological obstructions encoded in momentum-resolved 1D Hamiltonians,
and directly determined by the Wannier bands. We explicitly show the
emergent new lower bound of $\mathcal{G}$ in tilted 2D SSH model
and the anisotropic Wilson-Dirac  model. These results set new bounds
for physical quantities like the Wannier function spread and the optical
conductivity directly connected to the QMI. We further extend this
dimension-reduction framework to higher-order topology by defining
the quantum metric in the Wannier-band basis. The consequent Wannier-band
QMI $\mathcal{G}^{w}$ is directly connected to higher-order topology
indicated by the quadrupole moment in BBH model. Our results generalize
the conventional relation between topology and quantum metric by introducing
a dimension-reduction framework.

\section{Acknowledgments}

This work is financially supported by Innovation Program for Quantum
Science and Technology and the startup funding at HFNL (Grant No.
QD2022600001). F.B. is supported by is supported by National Natural
Science Foundation of China (Grants No.12504049), Guangdong Province
Introduced Innovative R\&D Team Program (Grant No. 2023QN10X136),
Guangdong Basic and Applied Basic Research Foundation No. 2024A1515010430
and 2023A1515140008). Y.Q. and J.L. were supported by the National
Natural Science Foundation of China under Grants No. 92265201 and
No. 12574176, and the Innovation Program for Quantum Science and Technology
under Project No. 2021ZD0302704.

\section{Data availability}

No data were created or analyzed in this study.

\appendix

\section{Energy spectrum and the winding number of 2D SSH }

\begin{figure}
\includegraphics[width=1\linewidth]{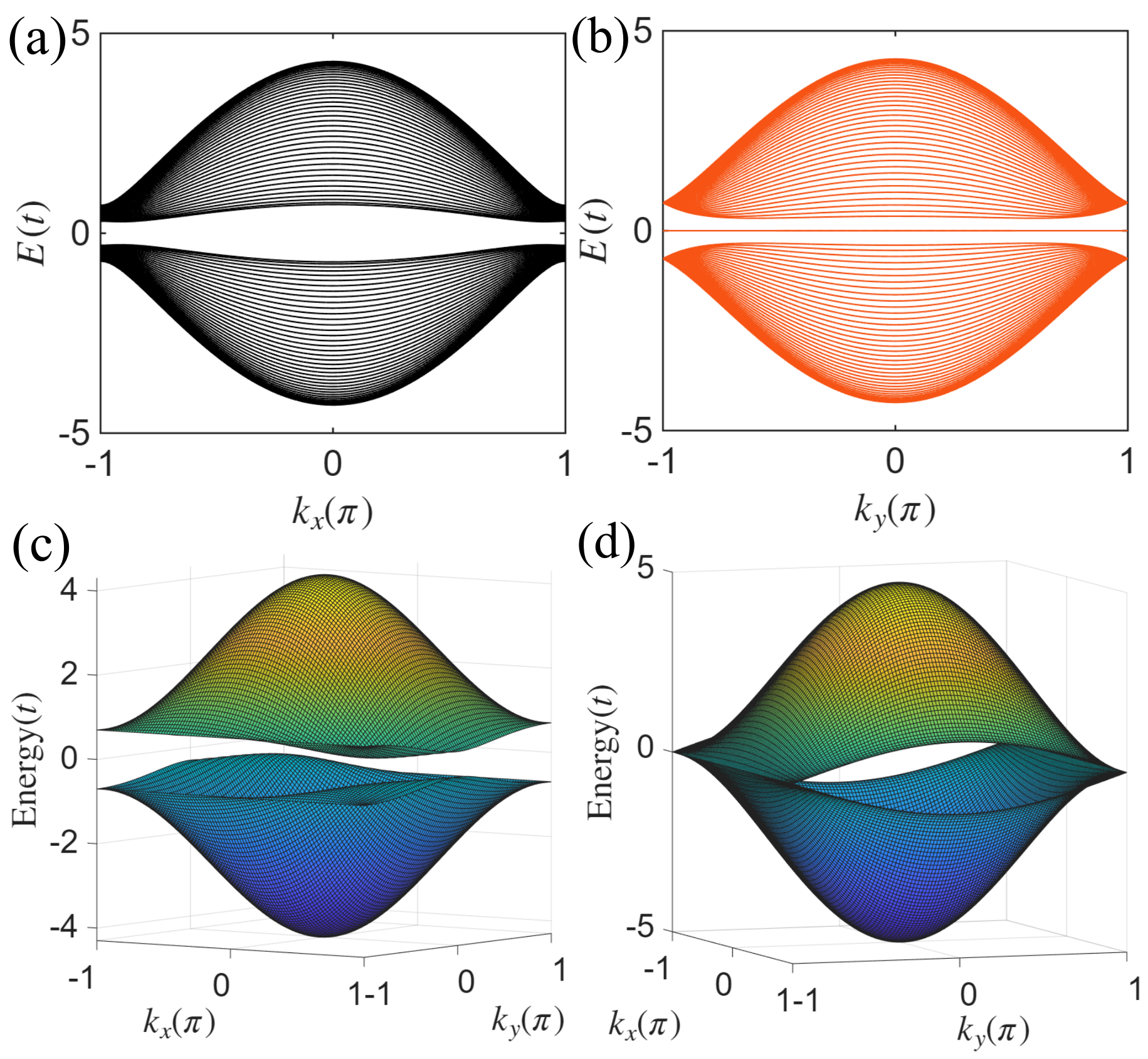}

\caption{Energy spectrum of the tilted 2D SSH model. (a) The energy spectrum
on a ribbon along $x$ direction. (b) The energy spectrum on a ribbon
along $y$ direction. There are zero-energy edge modes in the gap.
(c) The band structure for insulating phase. For (a-c), we take the
parameters $t_{1}=0.8t$ and $t_{2}=1.5t$, which set the system in
the WTP B. (d) The band structure for nodal-line semi-metal phase
at $t_{1}=t_{2}=1.5t$. \label{fig:Ribbonspectrum}}
\end{figure}

Here we present the energy spectrum in the WTP B as an example. In
this phase, the bulk band is gapped, as shown in Fig.\ \ref{fig:Ribbonspectrum}(c).
The energy spectra along different directions are shown in Fig.\ \ref{fig:Ribbonspectrum}(a)
and \ref{fig:Ribbonspectrum}(b). Due to the anisotropic topological
properties, zero-energy edge modes only appear for the ribbon along
$y$ direction. These edge modes correspond to quantized Wannier bands
in Fig.\ \ref{fig:model I}(d), and are protected by chiral symmetry.
For the WTP A, the zero-energy modes will appear for ribbons along
both $x$ and $y$ directions.

\section{QMI in the nodal-line phase}

For simplicity, we define $q({\bf k})=d_{x}+id_{y}=(1+e^{ik_{y}})(1+\tau e^{ik_{x}}\LyXZeroWidthSpace)$
in the Hamiltonian. It is seen that the nodal-line appears at $k_{y}=\pi$
for $q({\bf k})=0$. To regularize the occupied-band quantum metric,
we add a small mass term, $\ensuremath{H(\mathbf{k})=d_{x}\sigma_{x}+d_{y}\sigma_{y}+\Delta\sigma_{z}}.$
Near the nodal line, we take $k_{y}=\pi+\eta$ with $\eta\ll1$. Thus
$1+e^{ik_{y}}=1-e^{i\eta}\approx-i\eta$, and $q({\bf k})\approx-i\eta(1+\tau e^{ik_{x}})$.
Defining $\ensuremath{v(k_{x})=\left|1+\tau e^{ik_{x}}\right|^{2}=1+\tau^{2}+2\tau\cos k_{x}},$
we have $\ensuremath{d_{x}^{2}+d_{y}^{2}\simeq\eta^{2}v(k_{x})}$
and $\ensuremath{|\mathbf{d}|^{2}\simeq\eta^{2}v(k_{x})+\Delta^{2}}.$
From the two-band quantum metric formula, the leading divergent contribution
takes $\ensuremath{g_{yy}\simeq\frac{1}{4}\frac{\Delta^{2}v\left(k_{x}\right)}{\left[\eta^{2}v\left(k_{x}\right)+\Delta^{2}\right]^{2}}}$,
which is consistent with the result from Ref. \citep{Yugo24prx}.
Thus at the nodal line, $\ensuremath{g_{yy}\left(k_{y}=\pi\right)\simeq\frac{v\left(k_{x}\right)}{4\Delta^{2}}}$,
showing a divergent behavior in the local quantum metric. This divergent
behavior exists near the whole nodal-line with a width scales as $\delta\eta\sim|\Delta|$.
It is quite different from the nodal point case where the quantum
metric diverges only at single point. Therefore, the quantum metric
integral is estimated as $\ensuremath{\int d\eta g_{yy}\sim\frac{1}{|\Delta|}}.$

\section{Wilson loops and Wannier bands}

Here we present the projection method to get the Wannier bands \citep{LiCA20prb}.
The Wilson loop operator along $x$ direction is constructed as

\begin{equation}
\hat{P}_{x,{\bf k}}=P_{N_{x}\delta k_{x}+k_{x}}P_{(N_{x}-1)\delta k_{x}+k_{x}}\cdots P_{\delta k_{x}+k_{x}}P_{k_{x}},\label{eq:Wilsonloop}
\end{equation}
where each projection operator is defined as $P_{m\delta k_{x}+k_{x}}\equiv\sum_{n\in N_{\mathrm{occ}}}|u_{k_{y},m\delta k_{x}+k_{x}}^{n}\rangle\langle u_{k_{y},m\delta k_{x}+k_{x}}^{n}|$
with $|u_{k_{y},m\delta k_{x}+k_{x}}^{n}\rangle$ being the $n$-th
eigen state of occupied bands at point $(k_{y},m\delta k_{x}+k_{x})$,
and $m$ is an integer taking values from $\{1,2,\cdots,N_{y}\}$.
The projection method avoids the arbitrary phase problem in numerical
realizations. Here $N_{x}$ is the number of unit cells, $n$ is the
band index, and $N_{\mathrm{occ}}$ is the number of occupied bands.
Note that $\hat{P}_{x,{\bf k}}$ has dimension of $N$ with $N$ being
the total bands number. After projection onto the occupied bands at
base point ${\bf k}=(k_{x},k_{y})$, there is $N_{\mathrm{occ}}\times N_{\mathrm{occ}}$
matrix $\mathcal{W}_{x,{\bf k}}$ that defines a Wannier Hamiltonian
$H_{\mathcal{W}_{x}}({\bf k})$ from the relation $\mathcal{W}_{x,{\bf k}}=\exp[iH_{\mathcal{W}_{x}}({\bf k})]$.
The eigen values of $H_{\mathcal{W}_{x}}({\bf k})$ give the Wannier
bands $2\pi\nu_{x}(k_{y})$ associated with eigenstates $|\nu_{x,{\bf k}}^{j}\rangle$,
$j\in\{1,2,\cdots,N_{\mathrm{occ}}\}$. Similar procedure is applicable
for the construction of $\mathcal{W}_{y,{\bf k}}$.

\section{Wannier bands as odd function in Anisotropic Wilson-Dirac model}

In our calculations for model II of anisotropic Wilson-Dirac model,
we find that the Wannier bands satisfies $\nu_{x}(k_{y})+\nu_{x}(-k_{y})=0$
mod 1. Here we prove this relation as follows. The Hamiltonian in
Eq\textcolor{black}{.\ \eqref{eq:Model II}} satisfies 

\begin{alignat}{1}
H_{2}(k_{x},-k_{y}) & =H_{2}^{*}(k_{x},k_{y}).
\end{alignat}
Therefore, if $|u_{-}(k_{x},k_{y})\rangle$ is the occupied eigenstate
of $H_{2}(k_{x},k_{y})$, $|u_{-}(k_{x},k_{y})\rangle^{*}$ becomes
the eigenstate of $H_{2}(k_{x},-k_{y})$. Thus, we may choose the
gauge

\begin{alignat}{1}
|u_{-}(k_{x},-k_{y})\rangle & =|u_{-}(k_{x},k_{y})\rangle^{*}.
\end{alignat}
The Berry connection becomes 
\begin{alignat}{1}
A_{x}(k_{x},-k_{y}) & =i\langle u_{-}^{*}(k_{x},k_{y})|\partial_{k_{x}}u_{-}^{*}(k_{x},k_{y})\rangle\nonumber \\
 & =i\left(\langle u_{-}(k_{x},k_{y})|\partial_{k_{x}}u_{-}(k_{x},k_{y})\rangle\right)^{*}\nonumber \\
 & =-i\langle u_{-}(k_{x},k_{y})|\partial_{k_{x}}u_{-}(k_{x},k_{y})\rangle\nonumber \\
 & =-A_{x}(k_{x},k_{y}).
\end{alignat}
From the relation $\nu_{x}(k_{y})=\frac{1}{2\pi}\int dk_{x}A_{x}(k_{x},k_{y})$,
we have 
\begin{equation}
\nu_{x}(k_{y})+\nu_{x}(-k_{y})=0\ \mathrm{mod}\ 1.
\end{equation}
Similar results can be obtained for the Wannier band $\nu_{y}(k_{x})$. 

\section{Quantum metric inequality with multiple bands}

For the multi-band Hamiltonian with chiral symmetry, we can write
the Hamiltonian as

\begin{align}
H(k) & =\left(\begin{array}{cc}
0 & q(k)\\
q^{\dagger}(k) & 0
\end{array}\right).
\end{align}
After spectral flattening, the projection operator is obtained as
$P=\frac{1}{2}(I-H)$. Thus the quantum metric is 
\begin{align}
g_{xx}(k) & =\frac{1}{2}\mathrm{Tr}[\partial_{k}P\partial_{k}P]\nonumber \\
 & =\frac{1}{4}\mathrm{Tr}[\partial_{k}q\partial_{k}q^{\dagger}],
\end{align}
where we have used $\partial_{k}(qq^{\dagger})=0$. Note that the
winding number is defined as $w=\frac{1}{2\pi i}\int_{0}^{2\pi}\mathrm{Tr}\left[q^{\dagger}\partial_{k}q(k)\right]dk$.
Using the Cauchy-Schwartz inequality, we have

\begin{align}
\begin{aligned}\left(\int_{0}^{2\pi}-i\mathrm{Tr}\left[q^{\dagger}\partial_{k}q\right]dk\right)^{2} & \leq2\pi\int_{0}^{2\pi}-\left(\mathrm{Tr}\left[q^{\dagger}\partial_{k}q\right]\right)^{2}\\
 & =2\pi\int_{0}^{2\pi}\left|\mathrm{Tr}\left[q^{\dagger}\partial_{k}q\right]\right|^{2}.
\end{aligned}
\end{align}
Further noticing the matrix inequality $|\mathrm{Tr}A|^{2}\leq N_{\mathrm{occ}}\mathrm{Tr}\left(A^{\dagger}A\right)$
for a matrix $A$, it gives 

\begin{alignat}{1}
2\pi\int_{0}^{2\pi}\left|\mathrm{Tr}\left[q^{\dagger}\partial_{k}q\right]\right|^{2} & \leq2\pi N_{\mathrm{occ}}\int_{0}^{2\pi}\mathrm{Tr}\left(\partial_{k}q\partial_{k}q^{\dagger}\right)dk.
\end{alignat}
 Thus the QMI takes 
\begin{align}
\int_{0}^{2\pi}g_{xx}(k)dk & =\frac{1}{2}\int_{0}^{2\pi}\mathrm{Tr}\left[\partial_{k}P\partial_{k}P\right]dk\nonumber \\
 & =\frac{1}{4}\int_{0}^{2\pi}\mathrm{Tr}\left[\partial_{k}q\partial_{k}q^{\dagger}\right]dk\nonumber \\
 & \geq\frac{1}{4}\frac{1}{N_{\mathrm{occ}}}\frac{1}{2\pi}\left(\int_{0}^{2\pi}-i\mathrm{Tr}\left[q^{\dagger}\partial_{k}q\right]dk\right)^{2}\nonumber \\
 & =\frac{\pi}{2N_{\mathrm{occ}}}\left(\frac{1}{2\pi i}\int_{0}^{2\pi}\mathrm{Tr}\left[q^{\dagger}\partial_{k}q\right]dk\right)^{2}\nonumber \\
 & =\frac{\pi}{2N_{\mathrm{occ}}}w^{2},
\end{align}
where $N_{\mathrm{occ}}$ is the number of occupied bands and $w=\frac{1}{2\pi i}\int_{0}^{2\pi}\mathrm{Tr}\left[q^{\dagger}\partial_{k}q\right]dk$
is the winding number of the chiral-symmetric systems. 

\section{Quantum metric integral in SSH model}

Consider the SSH model with a Hamiltonian

\begin{alignat}{1}
H(k) & =h_{x}(k)\sigma_{x}+h_{y}(k)\sigma_{y}\nonumber \\
 & =(\lambda_{1}+\lambda_{2}\cos nk)\sigma_{x}+\lambda_{2}\sin(nk)\sigma_{y},
\end{alignat}
where $n$ is an integer. The corresponding eigenstates of $H(k)$
are $\psi_{\pm}=\frac{1}{\sqrt{2}}\binom{e^{-i\phi(k)}}{\pm1}$ with
$\tan\phi(k)=\frac{h_{y}}{h_{x}}$. The projector for occupied state
can be constructed as $P=|\psi_{-}\rangle\langle\psi_{-}|$. From
the QGT $g_{xx}=\frac{1}{2}\mathrm{Tr}[\partial_{k}P\partial_{k}P]$,
which gives 

\begin{alignat}{1}
\begin{aligned}g_{xx} & =\frac{1}{2}\int_{0}^{2\pi}(\partial_{k}\phi)^{2}dk\\
 & =\frac{1}{2}\int_{0}^{2\pi}n^{2}dk\left(\frac{1+t\cos nk}{1+t^{2}+2t\cos nk}\right)^{2}\\
 & =\frac{n}{2}\int_{0}^{2n\pi}d\theta\left(\frac{1+t\cos\theta}{1+t^{2}+2t\cos\theta}\right)^{2}\\
 & =\frac{n^{2}}{2}\int_{0}^{2\pi}d\theta\left(\frac{1+t\cos\theta}{1+t^{2}+2t\cos\theta}\right)^{2}.
\end{aligned}
\end{alignat}
Define the complex number $z=e^{i\theta}$, we have 

\begin{alignat}{1}
\frac{1}{1+tz} & =\frac{1}{1+te^{-i\theta}}=\frac{1+te^{i\theta}}{1+t^{2}+2t\cos\theta},
\end{alignat}
yielding
\begin{alignat}{1}
\frac{1+t\cos\theta}{1+t^{2}+2t\cos\theta} & =\mathrm{Re}\left(\frac{1}{1+te^{-i\theta}}\right).
\end{alignat}

Consider the topological nontrivial and trivial cases of the SSH model.
First, with $|t|<1$, the situation has $\frac{1}{1+te^{-i\theta}}=\sum_{k=0}^{+\infty}\left(-te^{-i\theta}\right)^{k}=\sum_{k}(-1)^{k}t^{k}e^{-i\theta k}$,
and $\mathrm{Re}\left(\frac{1}{1+te^{-i\theta}}\right)=\sum_{k}(-1)^{k}t^{k}\cos(\theta k)$.
Therefore, 

\begin{alignat}{1}
g_{xx} & =\frac{n^{2}}{2}\int_{0}^{2\pi}d\theta\left(\frac{1+t\cos\theta}{1+t^{2}+2t\cos\theta}\right)^{2}\nonumber \\
 & =\frac{n^{2}}{2}\int_{0}^{2\pi}d\theta\sum_{kk'}(-1)^{k}(-1)^{k^{\prime}}t^{k}t^{k^{\prime}}\cos(k\theta)\cos\left(k^{\prime}\theta\right),\nonumber \\
 & =n^{2}\frac{\pi}{2}\frac{1}{2}\left(\frac{2-t^{2}}{1-t^{2}}\right).
\end{alignat}
Second, with $|t|>1$, we have 
\begin{alignat}{1}
\begin{aligned}\frac{1}{1+te^{-i\theta}} & =\frac{\frac{1}{t}e^{i\theta}}{1+\frac{1}{t}e^{i\theta}}=\frac{1}{t}e^{i\theta}\sum_{k=0}^{+\infty}\left(-\frac{1}{t}\right)^{k}e^{i\theta k}\\
 & =\sum_{k=0}^{+\infty}(-1)^{k}\frac{1}{t^{k+1}}e^{i\theta(k+1)},
\end{aligned}
\end{alignat}
and $\mathrm{Re}\left(\frac{1}{1+te^{-i\theta}}\right)=\sum_{k=0}^{+\infty}(-1)^{k}\frac{1}{t^{k+1}}\cos[(k+1)\theta]$,
thus the QGT takes the form 

\begin{alignat}{1}
\begin{aligned} & \frac{n^{2}}{2}\int_{0}^{2\pi}d\theta\sum_{kk'}(-1)^{k+k^{\prime}}\frac{1}{t^{k+1}}\frac{1}{t^{k^{\prime}+1}}\cos[(k+1)\theta]\cos\left[\left(k^{\prime}+1\right)\theta\right]\\
 & \quad=\frac{n^{2}}{2}\sum_{kk'}(-1)^{k+k^{\prime}}\frac{1}{t^{2+k+k^{\prime}}}\delta_{k,k^{\prime}}\pi\\
 & \quad=\frac{n^{2}}{2}\pi\sum_{k=0}^{+\infty}\frac{1}{t^{2+2k}}=\frac{n^{2}}{2}\pi\frac{t^{-2}}{1-t^{-2}}=\frac{1}{t^{2}-1}\frac{n^{2}\pi}{2}.
\end{aligned}
\end{alignat}
Therefore, the QMI is 

\begin{alignat}{1}
\int_{0}^{2\pi}g_{xx}(k)dk & =\begin{cases}
n^{2}\frac{\pi}{2}\frac{1}{2}\left(1+\frac{1}{1-t^{2}}\right) & |t|<1,\\
n^{2}\frac{\pi}{2}\frac{1}{2}\left(\frac{1}{t^{2}-1}\right) & |t|>1.
\end{cases}\label{eq:QMI_SSH}
\end{alignat}
Note that the integer $n$ is the winding number of the SSH model.

From Eq\textcolor{black}{.\ \eqref{eq:QMI_SSH}, it is clear that
there are two cases for distinct topological phases. For the topological
nontrivial region with $|t|<1$, the QMI $\int_{0}^{2\pi}g_{xx}(k)dk$
is a monotonically increasing function of $t^{2}$. Therefore, it
has the minimal at $t=0$, which gives the value $\int_{0}^{2\pi}g_{xx}(k)dk=\frac{n^{2}\pi}{2}.$
This is consistent with the previous results $\int_{0}^{2\pi}g_{xx}(k)dk\geq\frac{n^{2}\pi}{2N}$
with $N=1$ at the saturation condition. The QMI is thus lowered bounded
by $\frac{n^{2}\pi}{2}$ with $n$ the winding number. For the topological
trivial region with $|t|>1$, the QMI $\int_{0}^{2\pi}g_{xx}(k)dk$
is a monotonically decreasing function of $t^{2}$. As $t^{2}\rightarrow\infty$,
the QMI $\int_{0}^{2\pi}g_{xx}(k)dk\rightarrow0$, which only has
trivial zero bound. Notably, at the phase transition point $|t|=1$
where the band gap closes, the QMI diverges. }

\section{Quantum metric of Dirac fermions in 2D}

Here we present the quantum metric for graphene. Consider the Hamiltonian
$H({\bf k})=k_{x}\sigma_{x}+k_{y}\sigma_{y}$, Take the lower eigen
state $|\psi\rangle=\frac{1}{\sqrt{2}}\left(\begin{array}{c}
e^{-i\phi}\\
-1
\end{array}\right),$ where $\phi\equiv\arctan\frac{k_{x}}{k_{y}}$. Thus in the polar
coordinates, the quantum metric components can be obtained as 
\begin{alignat}{1}
g_{kk}=0,g_{k\phi}=g_{\phi k}=0, & g_{\phi\phi}=\frac{1}{4}.
\end{alignat}
Thus the quantum distance element $ds^{2}=g_{\mu\nu}dx^{\mu}dx^{\nu}$
leads to $ds=\frac{1}{2}d\phi$. Thus the quantum distance is $\int ds=\int\frac{d\phi}{2}=\pi$.
Considering $\phi=\arctan(k_{y}/k_{x})$, it gives $d\phi=\frac{-k_{y}dk_{x}}{k^{2}}+\frac{k_{x}dk_{y}}{k^{2}}.$
It thus yields 
\begin{alignat}{1}
ds^{2} & =\frac{1}{4}d\phi^{2}=\frac{1}{4}\left(\frac{-k_{y}dk_{x}}{k^{2}}+\frac{k_{x}dk_{y}}{k^{2}}\right)^{2}\nonumber \\
 & =\frac{1}{4k^{4}}(k_{y}^{2}dk_{x}^{2}-2k_{x}k_{y}dk_{x}dk_{y}+k_{x}^{2}dk_{y}^{2}),
\end{alignat}
which gives the quantum metric in the Cartesian coordinate

\begin{alignat}{1}
g_{\mu\nu}({\bf k}) & =\frac{1}{4k^{4}}\left(\begin{array}{cc}
k_{y}^{2} & -k_{x}k_{y}\\
-k_{x}k_{y} & k_{x}^{2}
\end{array}\right).
\end{alignat}
At the Dirac point, the quantum metric will diverge and the it reflects
the singularities of the Dirac cone. 

\bibliographystyle{apsrev4-2-etal-title-url-first}

\end{document}